\RequirePackage{lineno}
\documentclass[aps,twocolumn,superscriptaddress,preprintnumbers,amsmath,amssymb]{revtex4}
\usepackage{graphicx,color,dcolumn,booktabs,bm}
\usepackage{longtable,lscape}
\usepackage{amsmath}
\usepackage{indentfirst}
\usepackage{epsfig}
\usepackage{feynmf}   
\usepackage{slashed}  
\usepackage{cases}
\usepackage{color}
\usepackage{multirow}
\usepackage{graphicx,color,dcolumn,booktabs,bm}
\usepackage{verbatim}
\usepackage[colorlinks,linkcolor=blue,anchorcolor=green,citecolor=blue]{hyperref}
\usepackage{mathrsfs}
\usepackage{rotating}
\usepackage{threeparttable}
\usepackage{anyfontsize}
\usepackage{orcidlink}

\newcommand{\BR}{{\cal B}}

\begin{document}

\title{\quad\\[1.0cm] Observation of $e^+e^-\to\omega\chi_{bJ}(1P)$ and search for $X_b \to \omega\Upsilon(1S)$ at $\sqrt{s}$ near 10.75 GeV}

  \author{I.~Adachi\,\orcidlink{0000-0003-2287-0173}} 
  \author{L.~Aggarwal\,\orcidlink{0000-0002-0909-7537}} 
  \author{H.~Ahmed\,\orcidlink{0000-0003-3976-7498}} 
  \author{H.~Aihara\,\orcidlink{0000-0002-1907-5964}} 
  \author{N.~Akopov\,\orcidlink{0000-0002-4425-2096}} 
  \author{A.~Aloisio\,\orcidlink{0000-0002-3883-6693}} 
  \author{N.~Anh~Ky\,\orcidlink{0000-0003-0471-197X}} 
  \author{D.~M.~Asner\,\orcidlink{0000-0002-1586-5790}} 
  \author{T.~Aushev\,\orcidlink{0000-0002-6347-7055}} 
  \author{V.~Aushev\,\orcidlink{0000-0002-8588-5308}} 
% ...
\collaboration{The Belle II Collaboration}

\begin{abstract}
We study the processes $e^+e^-\to\omega\chi_{bJ}(1P)$ ($J$ = 0, 1, or 2) using samples at center-of-mass energies $\sqrt{s}$ = 10.701, 10.745, and 10.805 GeV, corresponding to 1.6, 9.8, and 4.7 fb$^{-1}$ of integrated luminosity, respectively.
These data were collected with the Belle II detector during special operations of the SuperKEKB collider above the $\Upsilon(4S)$ resonance.
We report the first observation of $\omega\chi_{bJ}(1P)$ signals at $\sqrt{s}$ = 10.745 GeV. By combining Belle II data with Belle results at $\sqrt{s}$ = 10.867 GeV, we find energy dependencies of the Born cross sections for $e^+e^-\to \omega\chi_{b1,b2}(1P)$ to be consistent with the shape of the $\Upsilon(10753)$ state.
These data indicate that the internal structures of the $\Upsilon(10753)$ and $\Upsilon(10860)$ states may differ.
Including data at $\sqrt{s}$ = 10.653 GeV, we also search for the bottomonium equivalent of the $X(3872)$ state decaying into $\omega\Upsilon(1S)$. No significant signal is observed for masses between 10.45 and 10.65 GeV/$c^2$.

\end{abstract}

\maketitle

Heavy quarkonium spectroscopy offers multiple opportunities for insight on the nonperturbative behavior of quantum chromodynamics~\cite{1534,2981,074027}.
Four bottomonium-like vector states have been identified above the $B\bar B$ threshold~\cite{PDG}. The $\Upsilon(10753)$, observed in $e^+e^- \to \pi^+\pi^-\Upsilon(nS)$ ($n$ = 1, 2, 3) by Belle~\cite{220} and in fits to the $e^+e^-\to b\bar b$ cross sections at energies $\sqrt{s}$ from 10.6 to 11.2 GeV~\cite{083001}, is particularly interesting.

The $\Upsilon(10753)$ has been interpreted as a conventional bottomonium~\cite{074007,034036,59,014020,014036,357,04049,135340,11915,103845}, hybrid~\cite{034019,1}, or tetraquark state~\cite{135217,074507,11475,123102}.
Interpretations as an admixture of the conventional $4S$ and $3D$ states predict comparable branching fractions of $10^{-3}$ for $\Upsilon(10753)\to \pi^+\pi^-\Upsilon(nS)$~\cite{074007} and $\Upsilon(10753)\to \omega\chi_{bJ}$~\cite{034036}, where $\chi_{bJ}$ denotes $\chi_{bJ}(1P)$ throughout.
In this interpretation, the branching fraction for $\Upsilon(10753)\to\omega\chi_{b1}$ is expected to be about 1/5 of that for $\Upsilon(10753)\to\omega\chi_{b2}$~\cite{034036}.
In addition, the process $e^+e^- \to \gamma X_b, X_b \to \omega \Upsilon(1S)$, which shares the same final states as $\Upsilon(10753) \to \omega \chi_{bJ}$, provides access to the $X_b$.
The $X_b$ is the posited bottomonium counterpart of the $X(3872)$~\cite{262001} ($\chi_{c1}(3872)$~\cite{PDG}). 
Its existence has been predicted in molecular~\cite{525,054007,153,053001,238,243,100,1600,3063,034005,122001} and tetraquark models~\cite{214,567,28,014005}, with masses close to the $B\bar B^*$ threshold~\cite{525,054007,153,053001,238,243,100,1600,3063,034005,122001}, or in the 10 to 11 GeV/$c^2$ range~\cite{214,567,28,014005}, respectively.
Unlike in $X(3872)$ decays, due to negligible isospin-breaking effects, the $X_b$ may decay preferentially into $\pi^+\pi^-\pi^0\Upsilon(1S)$ instead of $\pi^+\pi^-\Upsilon(1S)$~\cite{1600,3063,034005,122001}.
Searches for an $X_b$ state in $\pi^+\pi^-\Upsilon(1S)$ by CMS and ATLAS~\cite{57,199}, and
in $e^+e^- \to \gamma X_b,X_b\to \pi^+\pi^-\pi^0\Upsilon(1S)$ near the $\Upsilon(10860)$ (known as the $\Upsilon(5S)$) energy by Belle~\cite{142001}, yielded null results.

In this Letter we test the nature of the $\Upsilon(10753)$ using measurements of Born cross sections for the $e^+e^-\to\omega\chi_{bJ}$ ($J$ = 0, 1, 2) processes at $\sqrt{s}$ between 10.701 and 10.805 GeV with $\omega\to\pi^+\pi^-\pi^0$, $\chi_{bJ}\to \gamma\Upsilon(1S)$, and $\Upsilon(1S)\to \ell^+\ell^-$ ($\ell$ = $e$ or $\mu$). We also search for the $X_b\to \omega \Upsilon(1S)$ process in $e^+e^-\to \gamma X_b$ at $\sqrt{s}$ between 10.653 and 10.805 GeV.
After event selection, fits to the
two-dimensional invariant mass distributions for $\omega$ and $\chi_{bJ}$ candidates
are used to search for $e^+e^-\to\omega \chi_{bJ}$ enhancements.
Observed event yields as functions of collision energy are used to extract cross sections, which, combined with previous Belle measurements at higher energy, determine the energy dependence. Finally, $M(\omega \Upsilon(1S))$ invariant mass distributions
are fitted to search for a $X_b$ signal.\par
These analyses are based on electron-positron collisions produced by the SuperKEKB collider~\cite{KEKB} at $\sqrt{s}$ = 10.653, 10.701, 10.745, and 10.805 GeV. Data samples, collected with the Belle II detector~\cite{Belle2}, correspond to integrated luminosities of 3.5, 1.6, 9.8, and 4.7 fb$^{-1}$~\cite{021001}, respectively.
The Belle II detector is a nearly $4\pi$ magnetic spectrometer surrounded by particle-identification detectors, an electromagnetic calorimeter, and muon and $K^0_L$ detectors~\cite{Belle2}. 
Simulated events are used for optimization of event selections and determination of reconstruction efficiencies and resolution functions.
We generate simulated signal events with {\sc evtgen}~\cite{152}.
Initial-state radiation (ISR) at next-to-leading order accuracy in quantum electrodynamics is simulated with {\sc phokhara}~\cite{71}. Angular distributions for the two-body decays in $e^+e^-\to \omega\chi_{bJ}$ and $\gamma X_b$ are generated according to a phase-space model. 
Simulated events are generated with varying $X_b$ masses from 10.45 to 10.65 GeV/$c^2$~\cite{525,054007,153,053001,238,243,100,1600,3063,034005,122001}.
The detector response is simulated with the Geant4 package~\cite{250}. Reconstruction of events from simulated and collision data is performed with the Belle II analysis software framework~\cite{31}.

Events are selected online by a trigger~\cite{1807} using central drift chamber and electromagnetic calorimeter information. 
In the offline analyses, all charged-particle trajectories (tracks) are required to originate from the vicinity of the interaction point.
We require four or five tracks to reduce backgrounds while allowing for increased efficiency for signal events with an additional misreconstructed track.
The identifications of pions, electrons, and muons are based on likelihood information from subdetectors~\cite{123C01}.
The pions are identified with 90\% efficiency and 8\% kaon contamination.
At least one of the leptons is identified with 95\% efficiency for electrons and 90\% efficiency for muons.
To reduce the effects of bremsstrahlung and final-state radiation, photons within a 50 mrad cone of the initial electron or positron direction are included in the calculation of the particle four-momentum.
The $\Upsilon(1S)$ signal regions are selected as 9.25 $<$ $M(e^+e^-)$ $<$ 9.58 GeV/$c^2$ and 9.34 $<$ $M(\mu^+\mu^-)$ $<$ 9.58 GeV/$c^2$.

Energy deposits in adjacent electromagnetic calorimeter crystals are treated as photon candidates if they are not associated with charged particles.
We require the energy of the photon from the $\chi_{bJ}$ decay to exceed 50 MeV.
Photons used to reconstruct $\pi^0$ candidates are required to have energies greater than 25, 25, and 40 MeV, when detected in the barrel, forward endcap, and backward endcap, respectively, and to satisfy $0.105<M(\gamma\gamma)<0.150$ GeV/$c^2$.
We perform mass-constrained fits for the $\Upsilon(1S)$ and $\pi^0$ candidates to improve momentum resolutions.

We perform kinematic fits to the $\pi^+\pi^-\pi^0\gamma\Upsilon(1S)$ combinations constraining their four-momenta to the initial 
$e^+e^-$ collision four-momentum.
An average of 1.18 candidates per event is found in data. 
In events with multiple candidates, only the candidate with the smallest fit $\chi^2$ is retained.
To avoid systematic effects from the modeling of the beam-energy spread, we do not use the momenta after the kinematic fit.

\begin{figure}[htbp]
\includegraphics[width=8.8cm]{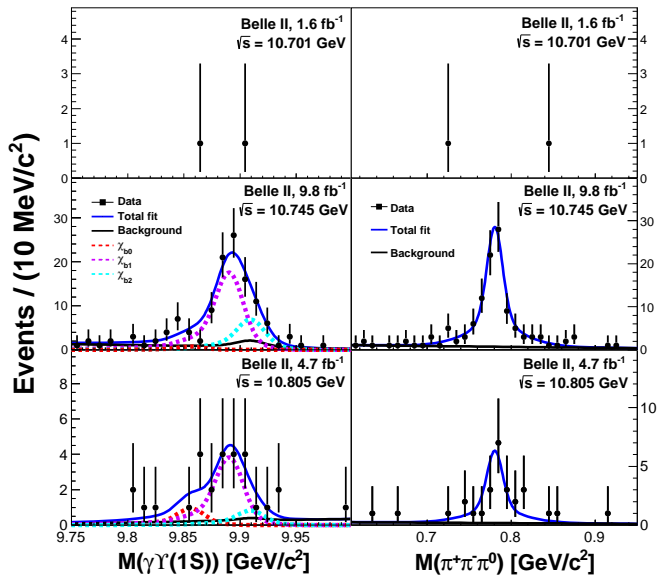}
\caption{Distributions of (left) $\gamma\Upsilon(1S)$ and (right) $\pi^+\pi^-\pi^0$ masses in data at $\sqrt{s}$ = 10.701, 10.745, and 10.805 GeV with fit results overlaid.}\label{fig1}
\end{figure}

Figure~\ref{fig1} shows distributions of $M(\gamma\Upsilon(1S))$ and $M(\pi^+\pi^-\pi^0)$ for events restricted to $9.75 < M(\gamma\Upsilon(1S)) < 10$ GeV/$c^2$ and $0.61 < M(\pi^+\pi^-\pi^0) < 0.95$ GeV/$c^2$, where signals are clearly visible at $\sqrt{s}$ = 10.745 and 10.805 GeV.

We perform a two-dimensional unbinned likelihood fit to the $M(\gamma\Upsilon(1S))$ versus $M(\pi^+\pi^-\pi^0)$ distributions for data collected at $\sqrt{s}$ = 10.745 and 10.805 GeV~\cite{SM1}. The fit function is a sum of four components: signals in $M(\gamma\Upsilon(1S))$ and $M(\pi^+\pi^-\pi^0)$ distributions, peaking background in the $M(\gamma\Upsilon(1S))$ distribution from $e^+e^-\to \pi^+ \pi^- \pi^0 \chi_{bJ}$, peaking background in the $M(\pi^+\pi^-\pi^0)$ distribution 
from non-$\chi_{bJ}$ backgrounds with a $\omega$,
and combinatorial background.
Each $\chi_{bJ}$ signal shape is described by a Crystal Ball function~\cite{CB} while the $\omega$ signal shape is described by a Breit-Wigner function (BW) convolved with a Gaussian function. The widths of the Crystal Ball and Gaussian functions are approximately 15 MeV/$c^2$ and 13 MeV/$c^2$, respectively.
Signal shape parameters are fixed from a fit to simulated signal events. 
A product of linear functions is used to describe the combinatorial
background.

Fit results are presented in Table~\ref{Tab1}. We find significant signals of 
$\chi_{b1}$ and $\chi_{b2}$ at $\sqrt{s}$ = 10.745 GeV. The significances
of the $\chi_{b0}$, $\chi_{b1}$, and $\chi_{b2}$ signals at 10.745 GeV and 10.805
GeV are 11$\sigma$ and 4.5$\sigma$, respectively. 
We report the first observation of $\omega\chi_{bJ}$ signals at $\sqrt{s}$ = 10.745 GeV.
These significances
are estimated using Wilks' theorem~\cite{significance}.
We compute 90\% Bayesian credibility upper limits on the yields assuming uniform priors (Table~\ref{Tab1}).
Systematic uncertainties in the Born cross-section measurements (discussed below) are approximately included by convolving the likelihood with a Gaussian function whose width equals the total systematic uncertainty. 

For data collected at $\sqrt{s}$ = 10.701 GeV, the signal yield is calculated as $N^{\rm sig}$ = max(0, $N^{\rm obs}$ $-$ $N^{b}$). The observed yield $N^{\rm obs}$ is obtained by counting events in the $\chi_{b0}$, $\chi_{b1}$, and $\chi_{b2}$ signal regions of 9.80 $<$ $M(\gamma\Upsilon(1S))$ $<$ 9.89 GeV/$c^2$, 9.84 $<$ $M(\gamma\Upsilon(1S))$ $<$ 9.93 GeV/$c^2$, and 9.86 $<$ $M(\gamma\Upsilon(1S))$ $<$ 9.95 GeV/$c^2$, respectively, in which about 95\% of signal candidates are retained; the background yields $N^{b}$ in the $\chi_{bJ}$ signal regions are obtained by scaling background events with the luminosity using $\sqrt{s}$ = 10.745 GeV data. The statistical uncertainty for $N^{\rm sig}$ is estimated using a Bayesian approach including background uncertainties~\cite{322}. The upper limit at 90\% Bayesian credibility on $N^{\rm sig}$ ($N^{\rm UL}$) is also determined with the same approach. Signal yields and upper limits are listed in Table~\ref{Tab1}.

The $e^+e^-\to \omega\chi_{bJ}$ Born cross section is calculated using 

\vspace{-0.8cm} 
\begin{equation} \label{eq:1}
\sigma_B(e^+e^-\to \omega\chi_{bJ}) = \frac{N^{\rm sig}\,  |1-\Pi|^2} {{\cal L}\,\varepsilon\, \BR_{\rm int}\,(1+\delta_{\rm ISR})},
\end{equation}
where ${\cal L}$ is the integrated luminosity, $\varepsilon$ is the reconstruction efficiency, $\BR_{\rm int}$ is the product of the branching fractions of the intermediate states, $|1-\Pi|^2$ is the vacuum polarization factor~\cite{083001,585}, and $(1+\delta_{\rm ISR})$ is the radiative-correction factor~\cite{isr,113009,2605}. In calculating the radiative-correction factor, we use the energy dependence of the Born cross sections measured in this work.
Values of the inputs, resulting Born cross sections, and their upper limits for non-significant signals are listed in Table~\ref{Tab1}. 

\renewcommand\arraystretch{1.2}
\begin{table*}[htbp!]
\caption{Measurements of $e^+e^-\to \omega\chi_{bJ}$ at $\sqrt{s}$ = 10.701, 10.745, and 10.805 GeV. $\Sigma$ is the signal significance; Syst is the systematic uncertainty. The first and second uncertainties (if available) indicate statistical and systematic contributions, respectively. The common systematic uncertainties for all energy points are 15.8\%, 9.4\%, and 9.3\% for $\omega\chi_{b0}$, $\omega\chi_{b1}$, and $\omega\chi_{b2}$, respectively.}
\centering
\vspace{0.2cm}
\label{Tab1}
\centering
\begin{tabular}{c c c c c c c c c c c}
\hline\hline
Channel & $\sqrt{s}$ (GeV) & $N^{\rm sig}$ & $N^{\rm UL}$ & $\Sigma(\sigma)$ & $\varepsilon$ & $|1-\Pi|^2$ & $1+\delta_{\rm ISR}$ & Syst (\%) & $\sigma_{B}$ (pb) & $\sigma^{\rm UL}_{B}$ (pb)  \\\hline
$e^+e^-\to \omega\chi_{b0}$ & 10.701& $0.0^{+1.1}_{-0.0}$ & 3.0 & - & 0.182 & 0.931 & 0.67 & 16.6 & $0.0^{+6.1}_{-0.0}$ & 16.6 \\
$e^+e^-\to \omega\chi_{b1}$ & & $0.0^{+2.1}_{-0.0}$ & 3.9 & - & 0.184 & 0.931 & 0.64 & 10.6 & $0.0^{+0.7}_{-0.0}$ & 1.2 \\
$e^+e^-\to \omega\chi_{b2}$ & & $0.1^{+2.2}_{-0.1}$ & 4.0 & - & 0.182 &0.931  & 0.62 & 10.6 & $0.1^{+1.4}_{-0.1}$ & 2.5 \\
$e^+e^-\to \omega\chi_{b0}$ & 10.745&$3.0^{+5.5}_{-4.7}$ & 12.0 & 0.5 & 0.183 & 0.931 & 0.65 & 25.9 & $2.8^{+5.1}_{-4.4}\pm0.7$ & 11.3 \\
$e^+e^-\to \omega\chi_{b1}$ & &$68.9^{+13.7}_{-13.5}$ & - & 5.9 & 0.183 & 0.931 & 0.65 & 12.7 & $3.6^{+0.7}_{-0.7}\pm0.5$ & - \\
$e^+e^-\to \omega\chi_{b2}$ & &$27.6^{+11.6}_{-10.0}$ & - & 3.1 & 0.184 & 0.931 & 0.65 & 14.5 & $2.8^{+1.2}_{-1.0}\pm0.4$ & - \\
$e^+e^-\to \omega\chi_{b0}$ & 10.805 &$3.6^{+3.8}_{-3.1}$ & 9.9 & 1.2 & 0.182 & 0.932 & 1.12 & 24.9 & $4.1^{+4.3}_{-3.5}\pm1.0$ & 11.4 \\
$e^+e^-\to \omega\chi_{b1}$ & &$15.0^{+6.8}_{-6.2}$ & 26.2 & 2.7 & 0.182 & 0.932 & 1.12 & 20.2 & $0.9^{+0.4}_{-0.4}\pm0.2$ & 1.7 \\
$e^+e^-\to \omega\chi_{b2}$ & &$3.3^{+5.3}_{-3.8}$ & 12.8 & 0.8 & 0.183 & 0.932 & 1.11 & 29.1 & $0.4^{+0.7}_{-0.5}\pm0.1$ & 1.6 \\
\hline\hline
\end{tabular}
\end{table*}

\begin{figure}[htbp]
\includegraphics[width=8.7cm]{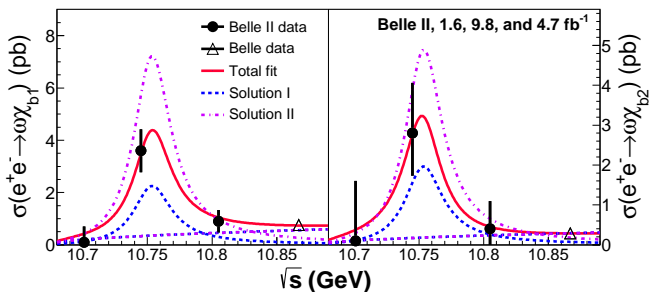}
\vspace{-0.5cm} 
\caption{Energy dependence of the Born cross sections for (left) $e^+e^- \to \omega\chi_{b1}$ and (right) $e^+e^- \to \omega\chi_{b2}$. Circles show the measurements reported here, triangles are the results of the Belle experiment~\cite{142001}. Error bars represent combined statistical and systematic uncertainties. Curves show the fit results and various components of the fit function.}\label{soluchib}
\end{figure}

Combining $\sigma_{B}(e^+e^-\to \omega\chi_{b1}~{\rm and}~\omega\chi_{b2})$ = $(0.76\pm0.16)~{\rm and}~(0.29\pm0.14)$ pb at $\sqrt{s} = 10.867$ GeV from Belle~\cite{142001} and those from this work, we show the Born cross sections for $e^+e^-\to \omega\chi_{b1}~{\rm and}~\omega\chi_{b2}$ as functions of collision energy in Fig.~\ref{soluchib}. 
We observe a strong enhancement of the cross section near 10.75 GeV.
We fit these distributions with a coherent sum of a two-body phase-space and a BW function~\cite{092007}

\vspace{-0.7cm} 
\begin{equation} \label{eq:BW}
|\sqrt{\Phi_2(\sqrt{s})}+
\frac{\sqrt{12\pi\Gamma_{ee}\BR_f\Gamma}} {s-M^2-iM\Gamma}\sqrt{ \frac{\Phi_2(\sqrt{s})}{\Phi_2(M)} } e^{i\phi}|^2,
\end{equation}where $M$ is the mass of the $\Upsilon(10753)$, 
$\Gamma$ ($\Gamma_{ee}$) is its total (electron) width,
$\BR_f$ is the branching fraction for the decay $\Upsilon(10753)\to\omega\chi_{b1}~{\rm or}~\omega\chi_{b2}$, $\Phi_2$ is the two-body phase-space factor, and $\phi$ is the relative phase between the amplitudes.~In the fits, the mass and total width in the BW function are fixed to 10752.7 MeV/$c^2$ and 35.5 MeV~\cite{220}, respectively.
There are two solutions for $\Gamma_{ee}\BR(\Upsilon(10753)\to\omega\chi_{b1}~{\rm and}~\omega\chi_{b2})$. One corresponds to constructive interference and yields $(0.63\pm0.39({\rm stat})\pm0.20({\rm syst}))$ and $(0.53\pm0.46({\rm stat})\pm0.15({\rm syst}))~{\rm eV}$ (Solution I).
The other corresponds to destructive interference and yields $(2.01\pm0.38({\rm stat})\pm0.76({\rm syst}))$ and $(1.32\pm0.44({\rm stat})\pm0.55({\rm syst}))~{\rm eV}$ (Solution II). 
The systematic uncertainties are discussed below.
The fit qualities for $\omega\chi_{b1}$ and $\omega\chi_{b2}$ are $\chi^2$/ndf = 0.5/1 and 0.1/1.
An alternative model with two interfering BW functions for the $\Upsilon(10753)$ and $\Upsilon(10860)$ states shows little interference and $\Gamma_{ee}\BR(\Upsilon(10753)\to\omega\chi_{b1}$ and $\omega\chi_{b2}) = (1.24\pm0.56({\rm stat}))$ and $(0.92\pm0.37({\rm stat}))$ eV (Solution I) and $(1.28\pm0.57({\rm stat}))$ and $(1.09\pm0.40({\rm stat}))$ eV (Solution II).
The fit qualities for $\omega\chi_{b1}$ and $\omega\chi_{b2}$ are $\chi^2$/ndf = 0.4/1 and 0.1/1.
We also test a single BW of $\Upsilon(10753)$ with a free mass and width to fit the energy dependence, and find a less satisfactory $\chi^2$/ndf of 12/6.

\begin{figure}[htbp]
\includegraphics[width=8.5cm]{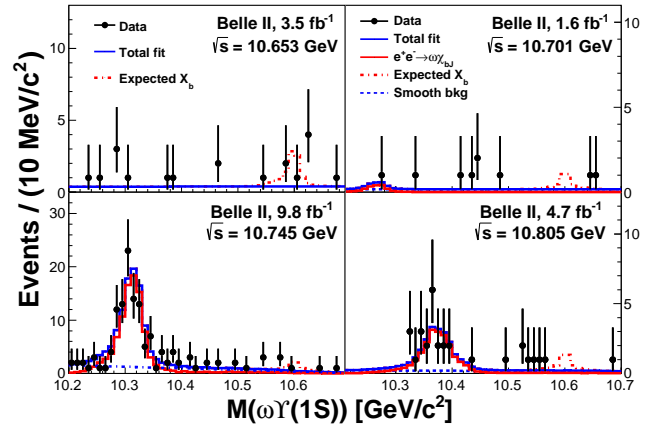}
\caption{Distributions of $\omega\Upsilon(1S)$ mass from data at $\sqrt{s}$ = 10.653, 10.701, 10.745, and 10.805 GeV.  The red dash-dotted histograms are from simulated events $e^+e^- \to \gamma X_b (\to \omega\Upsilon(1S))$ with the $X_b$ mass fixed at 10.6 GeV/$c^2$ and yields fixed at the upper limit values.}\label{dataXb}
\end{figure}

In addition, we search for the $X_b$ in $e^+e^-\to \gamma X_b$ with $X_b\to \omega\Upsilon(1S)$ at $\sqrt{s}$ = 10.653, 10.701, 10.745, and 10.805 GeV. 
Distributions of $M(\omega\Upsilon(1S))$ for events within 0.70 $<$ $M(\pi^+\pi^-\pi^0)$ $<$ 0.86 GeV/$c^2$ are shown in Fig.~\ref{dataXb}. Prominent reflections of $e^+e^-\to \omega\chi_{bJ}$ signals are observed, but no narrow structure as expected from a $X_b$ signal is found. We set upper limits at 90\% Bayesian credibility~\cite{322} with $M(\omega\Upsilon(1S))$ in the 100 MeV/$c^2$ interval centered at the test $X_b$ mass (95\% signal efficiency).
The observed yield is obtained by counting events, and 
the background yield is estimated using an unbinned maximum likelihood fit to the 
$M(\omega\Upsilon(1S))$ distribution. A straight line is used to describe the smooth 
background. 
At $\sqrt{s}$ = 10.701, 10.745, and 10.805 GeV, additional normalized shapes from simulations for each $\chi_{bJ}$ state are used to describe $e^+e^-\to\omega\chi_{bJ}$ signals.

We calculate the upper limits at 90\% Bayesian credibility on the products of the Born cross section for the $e^+e^-\to \gamma X_b$ process and the branching fraction for $X_b \to \omega\Upsilon(1S)$ ($\sigma^{\rm UL}_{B}(e^+e^-\to\gamma X_b)\BR(X_b \to \omega\Upsilon(1S)$) with varying $X_b$ masses from 10.45 to 10.65 GeV/$c^2$ using a similar relationship as Eq.~(\ref{eq:1}). The least stringent limits on signal yields are 10.0, 8.1, 8.1, 10.7 with $m(X_b)$ = 10.59, 10.45, 10.45, and 10.53 GeV/$c^2$ at $\sqrt{s}$ = 10.653, 10.701, 10.745, 10.805 GeV, respectively, with an efficiency of about 16\%. The values of $\sigma^{\rm UL}_{B}(e^+e^-\to\gamma X_b)\BR(X_b \to \omega\Upsilon(1S))$ range in (0.14 -- 0.55) pb, (0.25 -- 0.84) pb, (0.06 -- 0.14) pb, and (0.08 -- 0.37) pb, respectively~\cite{SM3}.

The individual sources of systematic uncertainties considered in the measurements of $\sigma_B(e^+e^-\to \omega\chi_{bJ})$ and $\sigma_B(e^+e^-\to \gamma X_b)\BR(X_b \to \omega\Upsilon(1S))$ are listed in the Supplemental Material~\cite{SM2} and the total systematic uncertainties are listed in Table~\ref{Tab1}.
Detection efficiency uncertainties include momentum-dependent tracking uncertainties (1.3\% per pion and 0.3\% per lepton, as derived from $\bar B^0\to D^{*+}(\to D^0\pi^+)\pi^-$ and $e^+e^-\to \tau^+\tau^-$), particle identification (1.1\% per pion, as derived from $D^{*+} \to D^0(\to K^-\pi^+)\pi^+$), lepton identification (0.4\% per electron and 0.7\% per muon, as derived from $J/\psi$ decays, Bhabha, dimuon, and two-photon processes), photon reconstruction (3.5\% per photon, as derived from $e^+e^-\to \gamma\mu^+\mu^-$), and $\pi^0$ reconstruction (4.8\% per $\pi^0$, as derived from $\eta\to\pi^0\pi^0\pi^0$).~The uncertainties on the branching fractions, 14.7\%, 7.4\%, and 7.3\% for $\chi_{b0}$, $\chi_{b1}$, and $\chi_{b2}$ decays,
are taken from Ref.~\cite{PDG}.
For $\sigma_{B}(e^+e^-\to \omega\chi_{bJ})$, we change the mass and width values of $\Upsilon(10753)$ by one standard deviation~\cite{220}, and the fit model from the coherent sum of a two-body phase-space and a BW function to two interfering BW functions for the $\Upsilon(10753)$ and $\Upsilon(10860)$. The differences on the radiative correction factors are 2.0\%, 5.1\%, and 13.7\% for $\sqrt{s}$ = 10.701, 10.745, and 10.805 GeV, respectively.
We assume $\sigma(e^+e^-\to\gamma X_b)$ to be proportional to $1/s^2$ to determine the radiative-correction factor for $e^+e^-\to \gamma X_b$. The maximum difference on the radiative-correction factor observed by using $1/s$ or $1/s^3$ is included as an uncertainty, which is less than 1\%. 
The uncertainty associated with the assumption of uniform distribution for the $\omega\chi_{bJ}$ (or $\gamma X_b$) particles is assessed by using an alternative signal efficiency based on a $1\pm{\rm cos}^2\theta$ distribution, where $\theta$ is the $\omega$ ($\gamma$) angle in $e^+e^-$ center-of-mass frame.
For $\sigma_B(e^+e^-\to \omega\chi_{bJ})$ at $\sqrt{s}$ = 10.745 and 10.805 GeV, we estimate the systematic uncertainties associated with the fit by changing the order of the background polynomial and the range of the fit, and comparing the fit results without a $\chi_{b0}$ component. The resulting systematic uncertainties are 16.3\%,~4.6\%,~8.2\% and 10.9\%,~8.9\%,~20.0\% for $\chi_{b0,b1,b2}$ decays at $\sqrt{s}$ = 10.745 and 10.805 GeV.
Studies in the control channel $e^+e^-\to \pi^+\pi^+\pi^-\pi^-\pi^0\pi^0$ show that a 1\% uncertainty needs to be included due to mismodeling in the trigger simulation.
We assume a beam-energy calibration uncertainty of 5 MeV and take the fractional differences in signal yields as uncertainties. These are 10.5\%,~2.5\%,~3.0\% and 6.5\%,~5.0\%,~12.2\% for $\chi_{b0,b1,b2}$ decays at $\sqrt{s}$ = 10.745 and 10.805 GeV. Belle II measures the luminosity at a 0.6\% precision~\cite{021001}.
Systematic uncertainties on the $\Gamma_{ee}\BR(\Upsilon(10753)\to\omega\chi_{b1}~{\rm and}~\omega\chi_{b2})$ determination come from 
variations of the $\Upsilon(10753)$ parameters and collision energies within their uncertainties.

In summary, we use a unique sample of electron-positron data collected at center-of-mass energies between 10.653 and 10.805 GeV to measure the Born cross sections for the $e^+e^-\to\omega\chi_{bJ}$ process and search for the $X_b$ state decaying into $\omega\Upsilon(1S)$.
A significant $\omega\chi_{b1}$ signal and evidence for the $e^+e^-\to\omega\chi_{b2}$ process at $\sqrt{s}$ = 10.745 GeV are found. The corresponding Born cross sections are $(3.6^{+0.7}_{-0.7}({\rm stat})\pm0.5({\rm syst}))$ pb and $(2.8^{+1.2}_{-1.0}({\rm stat})\pm0.4({\rm syst}))$ pb. We observe a strong enhancement of the cross section near 10.75 GeV whose energy dependence is consistent with the $\Upsilon(10753)$ state. We measure the products $\Gamma_{ee}\BR(\Upsilon(10753)\to\omega\chi_{b1}~{\rm and}~\omega\chi_{b2})$ to be in the range 0.20--2.9 and 0.05--2.0 eV.

The observed ratio $\sigma_B(e^+e^-\to \omega\chi_{b1})/\sigma_B(e^+e^-\to \omega\chi_{b2})$ = $1.3\pm0.6$ at $\sqrt{s}$ = 10.745 GeV,
where the statistical uncertainties and all the uncommon systematic uncertainties are included, contradicts the expectation for a pure $D-$wave bottomonium state of 15~\cite{172,comment}.
There is also a 1.8$\sigma$ difference with the prediction for a $S-D-$mixed state of 0.2~\cite{034036}.

This analysis shows that the $e^+e^-\to\omega\chi_{bJ}$ process observed near $\Upsilon(10860)$ by Belle could be due to the tail of the $\Upsilon(10753)$.
The large difference between the small value of the $\omega\chi_{bJ}$ to $\pi^+\pi^-\Upsilon(nS)$ production rate at the $\Upsilon(10860)$ and the value at the $\Upsilon(10753)$ may indicate different internal structures for these two states, which otherwise have the same quantum numbers and are only 110 MeV/$c^2$ apart.

No evidence of a $X_b$ signal is obtained for $X_b$ masses between 10.45 and 10.65 GeV/$c^2$, and upper limits at 90\% Bayesian credibility on the products of Born cross section for $e^+e^-\to \gamma X_b$ and branching fraction for $X_b \to \omega\Upsilon(1S)$ are set to be 0.55, 0.84, 0.14, and 0.37 pb at $\sqrt{s}$ = 10.653, 10.701, 10.745, and 10.805 GeV, respectively.

We warmly thank Fengkun Guo, Xiang Liu, Dianyong Chen, and Zhiyong Zhou for valuable and helpful discussions.~This work, based on data collected using the Belle II detector, which was built and commissioned prior to March 2019, was supported by
Science Committee of the Republic of Armenia Grant No.~20TTCG-1C010;
Australian Research Council and research Grants
No.~DE220100462,
No.~DP180102629,
No.~DP170102389,
No.~DP170102204,
No.~DP150103061,
No.~FT130100303,
No.~FT130100018,
and
No.~FT120100745;
Austrian Federal Ministry of Education, Science and Research,
Austrian Science Fund
No.~P~31361-N36
and
No.~J4625-N,
and
Horizon 2020 ERC Starting Grant No.~947006 ``InterLeptons'';
Natural Sciences and Engineering Research Council of Canada, Compute Canada and CANARIE;
Chinese Academy of Sciences and research Grant No.~QYZDJ-SSW-SLH011,
National Natural Science Foundation of China and research Grants
No.~11521505,
No.~11575017,
No.~11675166,
No.~11761141009,
No.~11705209,
and
No.~11975076,
LiaoNing Revitalization Talents Program under Contract No.~XLYC1807135,
Shanghai Municipal Science and Technology Committee under Contract No.~19ZR1403000,
Shanghai Pujiang Program under Grant No.~18PJ1401000,
and the CAS Center for Excellence in Particle Physics (CCEPP);
the Ministry of Education, Youth, and Sports of the Czech Republic under Contract No.~LTT17020 and
Charles University Grant No.~SVV 260448 and
the Czech Science Foundation Grant No.~22-18469S;
European Research Council, Seventh Framework PIEF-GA-2013-622527,
Horizon 2020 ERC-Advanced Grants No.~267104 and No.~884719,
Horizon 2020 ERC-Consolidator Grant No.~819127,
Horizon 2020 Marie Sklodowska-Curie Grant Agreement No.~700525 "NIOBE"
and
No.~101026516,
and
Horizon 2020 Marie Sklodowska-Curie RISE project JENNIFER2 Grant Agreement No.~822070 (European grants);
L'Institut National de Physique Nucl\'{e}aire et de Physique des Particules (IN2P3) du CNRS (France);
BMBF, DFG, HGF, MPG, and AvH Foundation (Germany);
Department of Atomic Energy under Project Identification No.~RTI 4002 and Department of Science and Technology (India);
Israel Science Foundation Grant No.~2476/17,
U.S.-Israel Binational Science Foundation Grant No.~2016113, and
Israel Ministry of Science Grant No.~3-16543;
Istituto Nazionale di Fisica Nucleare and the research grants BELLE2;
Japan Society for the Promotion of Science, Grant-in-Aid for Scientific Research Grants
No.~16H03968,
No.~16H03993,
No.~16H06492,
No.~16K05323,
No.~17H01133,
No.~17H05405,
No.~18K03621,
No.~18H03710,
No.~18H05226,
No.~19H00682, 
No.~22H00144,
No.~26220706,
and
No.~26400255,
the National Institute of Informatics, and Science Information NETwork 5 (SINET5), 
and
the Ministry of Education, Culture, Sports, Science, and Technology (MEXT) of Japan;  
National Research Foundation (NRF) of Korea Grants
No.~2016R1\-D1A1B\-02012900,
No.~2018R1\-A2B\-3003643,
No.~2018R1\-A6A1A\-06024970,
No.~2018R1\-D1A1B\-07047294,
No.~2019K1\-A3A7A\-09033840,
No.~2019R1\-I1A3A\-01058933,
and
No.~2022R1\-A2C\-1003993,
Radiation Science Research Institute,
Foreign Large-size Research Facility Application Supporting project,
the Global Science Experimental Data Hub Center of the Korea Institute of Science and Technology Information
and
KREONET/GLORIAD;
Universiti Malaya RU grant, Akademi Sains Malaysia, and Ministry of Education Malaysia;
Frontiers of Science Program Contracts
No.~FOINS-296,
No.~CB-221329,
No.~CB-236394,
No.~CB-254409,
and
No.~CB-180023, and No.~SEP-CINVESTAV research Grant No.~237 (Mexico);
the Polish Ministry of Science and Higher Education and the National Science Center;
the Ministry of Science and Higher Education of the Russian Federation,
Agreement No.~14.W03.31.0026, and
the HSE University Basic Research Program, Moscow;
University of Tabuk research Grants
No.~S-0256-1438 and No.~S-0280-1439 (Saudi Arabia);
Slovenian Research Agency and research Grants
No.~J1-9124
and
No.~P1-0135;
Agencia Estatal de Investigacion, Spain
Grant No.~RYC2020-029875-I
and
Generalitat Valenciana, Spain
Grant No.~CIDEGENT/2018/020
Ministry of Science and Technology and research Grants
No.~MOST106-2112-M-002-005-MY3
and
No.~MOST107-2119-M-002-035-MY3,
and the Ministry of Education (Taiwan);
Thailand Center of Excellence in Physics;
TUBITAK ULAKBIM (Turkey);
National Research Foundation of Ukraine, project No.~2020.02/0257,
and
Ministry of Education and Science of Ukraine;
the U.S. National Science Foundation and research Grants
No.~PHY-1913789 
and
No.~PHY-2111604, 
and the U.S. Department of Energy and research Awards
No.~DE-AC06-76RLO1830, 
No.~DE-SC0007983, 
No.~DE-SC0009824, 
No.~DE-SC0009973, 
No.~DE-SC0010007, 
No.~DE-SC0010073, 
No.~DE-SC0010118, 
No.~DE-SC0010504, 
No.~DE-SC0011784, 
No.~DE-SC0012704, 
No.~DE-SC0019230, 
No.~DE-SC0021274, 
No.~DE-SC0022350; 
and
the Vietnam Academy of Science and Technology (VAST) under Grant No.~DL0000.05/21-23.

These acknowledgements are not to be interpreted as an endorsement of any statement made
by any of our institutes, funding agencies, governments, or their representatives.

We thank the SuperKEKB group for the excellent operation of the accelerator and their special efforts to accomplish this center-of-mass energy scan;
the KEK cryogenics group for the efficient operation of the detector solenoid magnet;
the KEK computer group and the NII for on-site computing support and SINET6 network support;
and the raw-data centers at BNL, DESY, GridKa, IN2P3, INFN, and the University of Victoria for offsite computing support.


\begin{thebibliography}{**}
\bibitem{comment} In Ref.~\cite{172}, the ratio is given to the $\Upsilon(10860)$ decays under the assumption of $D-$wave. After discussing with the authors of Ref.~\cite{172}, such ratio is also applicable for the $\Upsilon(10753)$.
\end{thebibliography}
\end{document}